\documentclass[pre,showpacs,showkeys,preprintnumbers,amsmath,amssymb,superscriptaddress,twocolumn]{revtex4-2}
\usepackage{graphicx}
\usepackage{amssymb,amsfonts,amsmath}
\usepackage{epstopdf}
\usepackage{color}
\usepackage{bm}
\usepackage[colorlinks=true,linkcolor=blue,citecolor=red]{hyperref}

\def\I{{\bf I}}
\def\M{{\bf M}}

\begin{document}

\title{Improved boundary homogenization for a sphere with an absorbing cap of arbitrary size} 

\author{Denis~S.~Grebenkov}
 \email{denis.grebenkov@polytechnique.edu}

\affiliation{Laboratoire de Physique de la Mati\`{e}re Condens\'{e}e (UMR 7643), \\ 
CNRS -- Ecole Polytechnique, Institut Polytechnique de Paris, 91120 Palaiseau, France}

\date{Received: \today / Revised version: }

\begin{abstract}
Finding accurate approximations for the effective reactivity of a
structured spherical target with a circular absorbing patch of
arbitrary size is a long-standing problem in chemical physics.  In
this Communication, we reveal limitations of the empirical
approximation proposed in [J. Chem. Phys. 145, 214101 (2016)].  We
show that the original approximation fails at large patch surface
fractions $\sigma$ and propose a simple amendment.  The improved
approximation is validated against a semi-analytical solution and is
shown to be accurate over the entire range of $\sigma$ from $0$ to
$1$.  This approximation also determines the probability of reaction
on the patch and the capacitance of such a structured target.
\end{abstract}



\maketitle 

\begin{figure}[t!]
\begin{center}
\includegraphics[width=40mm]{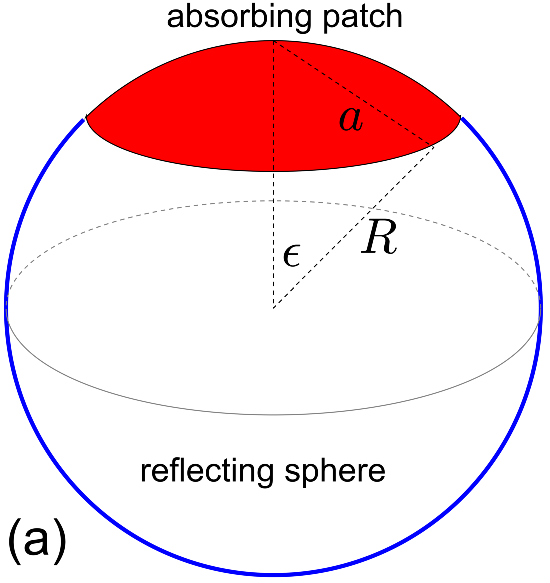} 
\includegraphics[width=88mm]{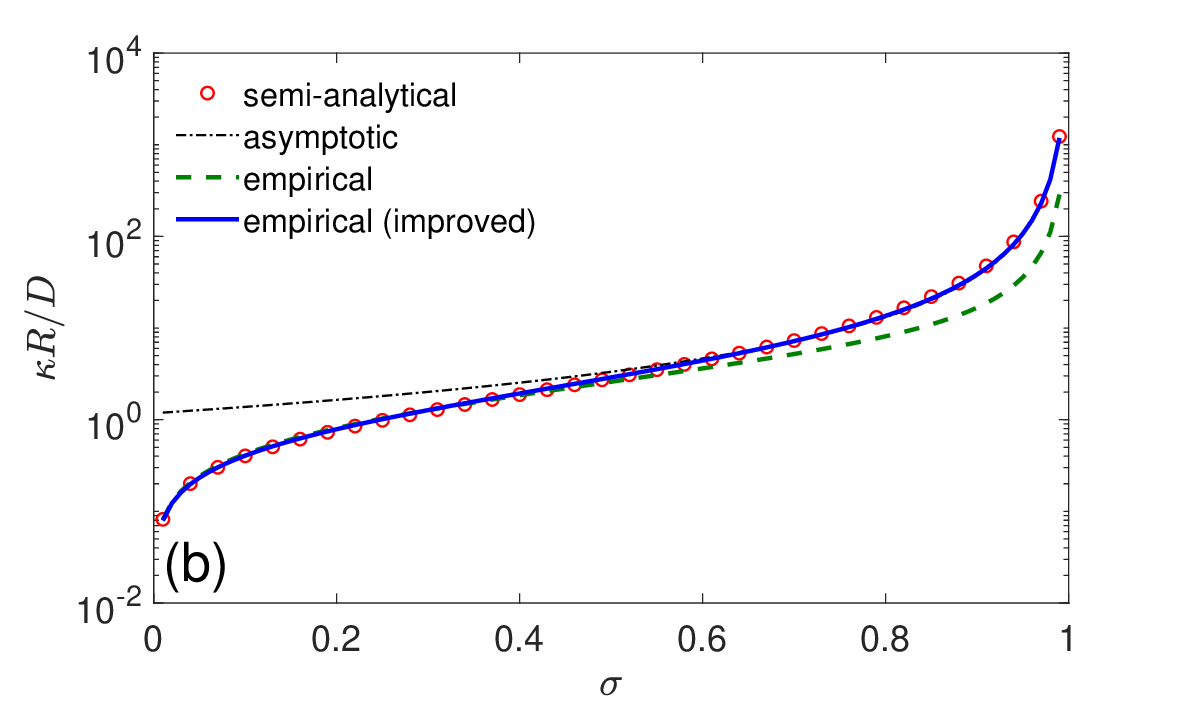} 
\includegraphics[width=88mm]{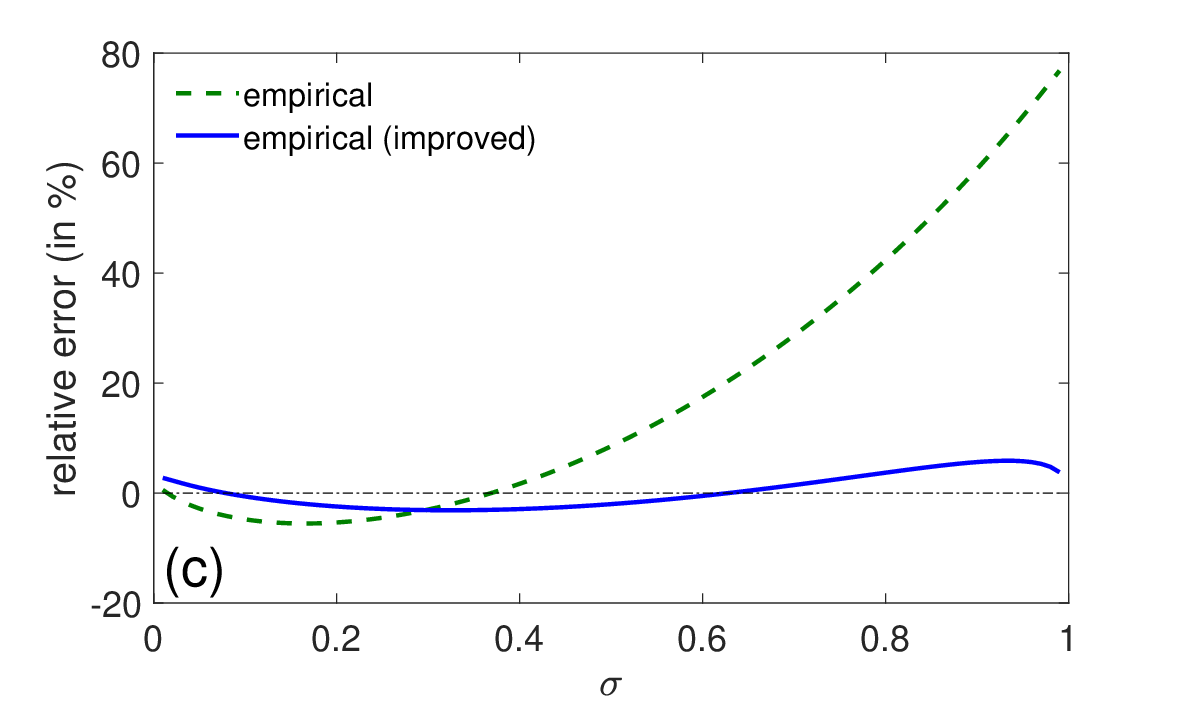} 
\end{center}
\caption{
{\bf (a)} A perfectly absorbing circular patch of radius $a$ located
on the North pole of the otherwise reflecting sphere of radius $R$.
{\bf (b)} Rescaled effective reactivity $\kappa R/D$ as a function of
the patch surface fraction $\sigma$.  Solid line shows the
semi-analytical solution (\ref{eq:Traytak}) obtained by truncating the
matrix $\M$ to the size $100 \times 100$, dash-dotted line indicates
the asymptotic formula (\ref{eq:asymptotic}), dashed line presents the
empirical approximation (\ref{eq:Dagdug}), while circles give the
improved empirical approximation (\ref{eq:DG}).  {\bf (c)} Relative
error of empirical approximations (\ref{eq:Dagdug}, \ref{eq:DG}) as
compared to the semi-analytical solution (\ref{eq:Traytak}).}
\label{fig:Dagdug}
\end{figure}

We revisit an emblematic problem of steady-state diffusion outside a
three-dimensional inert ball of radius $R$ with a perfectly absorbing
circular patch $\Gamma$ of arbitrary radius $a$ on its boundary
(Fig. \ref{fig:Dagdug}(a)).  Fixing a concentration $C_0$ at infinity,
the diffusive flux $J$ onto the absorbing patch can be expressed as $J
= \int\nolimits_{\Gamma} (D\partial_r c) dS$, where $D$ is the
diffusion coefficient.  Here the steady-state concentration
$c(r,\theta,\phi)$, written in spherical coordinates, satisfies the
Laplace equation, $\Delta c = 0$, with $c\to C_0$ as $r\to \infty$,
and mixed Dirichlet-Neumann boundary conditions on the sphere (at $r =
R$): $c = 0$ on the absorbing patch ($0 \leq \theta <
\epsilon$) and $\partial_r c = 0$ on the remaining reflecting boundary
($\epsilon \leq \theta \leq \pi$), with $\epsilon = 2\arcsin(a/(2R))$.
The effective reactivity $\kappa$ of such a structured target is then
defined by requiring that the diffusive flux $J$ is equal to the flux
of reacted particles on the homogenized spherical boundary of
reactivity $\kappa$ \cite{Collins49}
\begin{equation}
J = \frac{J_S}{1 + D/(\kappa R)}  \quad \Rightarrow \quad \frac{R}{D} \kappa = \frac{1}{J_S/J - 1} \,,
\end{equation}
where $J_S = 4\pi D R C_0$ is the Smoluchowski flux onto a perfectly
absorbing sphere of radius $R$.

As the above boundary value problem does not admit an explicit
analytical solution, various numerical schemes and approximations were
developed.  In particular, an empirical approximate formula for the
effective reactivity $\kappa$ was proposed in \cite{Dagdug16}:
\begin{equation} \label{eq:Dagdug}
\frac{R}{D} \kappa_{\rm emp} = \frac{2}{\pi} \sqrt{\sigma} \frac{1 + 2.6\sqrt{\sigma} - 3.2 \sigma^2}{(1-\sigma)^{3/2}} \,,
\end{equation}
where $\sigma = \pi a^2/(4\pi R^2)$ is the fraction of the sphere
covered by the absorbing patch.  This formula, which was constructed
from the numerical values of the reactivity estimated from Brownian
dynamics simulations, satisfies the well-known asymptotic behavior in
the small-$\sigma$ limit.  We aim at revealing and amending
limitations of this formula in the limit $\sigma \to 1$ when only a
small fraction of the sphere is not absorbing.

To inspect this limit, we recall that the effective reactivity
$\kappa$ is inversely proportional to the blockage coefficient of a
potential flow in a tube with a blocking disk
\cite{Martin20,Martin22,Skvortsov23}.  In the limit of a small blocking
disk, the approximation (32) from \cite{Martin20} yields the following
asymptotic behavior (see also the explicit relation (11.39b) from
\cite{Grebenkov24}): 
\begin{equation}  \label{eq:asymptotic}
\frac{R}{D} \kappa \approx \frac{R}{\sqrt{S}} \, \frac{3\pi^{3/2}}{4(1-\sigma)^{3/2}} = \frac{3\pi}{8} (1-\sigma)^{-3/2} \quad (\sigma\to 1),
\end{equation}
where $S = 4\pi R^2$ is the surface area.  This behavior also follows
from the asymptotic analysis of the steric factor in \cite{Dagdug17}.
In turn, the empirical formula (\ref{eq:Dagdug}) yields $\frac{R}{D}
\kappa_{\rm emp}(\sigma) \approx \tfrac{4}{5\pi} (1 - \sigma)^{-3/2}$,
with a different numerical prefactor.  This difference has a dramatic
effect onto the effective reactivity.  In particular, the relative
error of the empirical approximation, $1 - \kappa_{\rm
emp}(\sigma)/\kappa(\sigma)$, approaches $1 - 32/(15\pi^2)
\approx 0.78$ as $\sigma\to 1$.  In other words, the empirical formula
(\ref{eq:Dagdug}) significantly underestimates the effective
reactivity for large $\sigma$ that limits the range of its
applicability.

To illustrate these limitations, one has to solve the mixed boundary
value problem for the concentration and then to compute the flux $J$.
Traytak reformulated it in terms of dual series and found its formal
semi-analytical solution, from which the effective reactivity follows
as
\begin{equation}  \label{eq:Traytak}
\frac{R}{D}\kappa = (1/P-1)^{-1}, ~~ 
P = \frac{J}{J_S} = \sum\limits_{k=0}^\infty \bigl[(\I - \M)^{-1}\bigr]_{0,k} \M_{k,0} ,
\end{equation} 
where $\I$ is the identity matrix, and the elements of the
infinite-dimensional matrix $\M$ are 
\begin{equation}
\M_{l,m} = \frac{1}{2\pi(m+1)} \biggl(\frac{\sin((l+m+1)\epsilon)}{l+m+1} + \frac{\sin((l-m)\epsilon)}{l-m}\biggr),
\end{equation}
with $l,m = 0,1,2,\cdots$ (note that the second term should be
understood as $\epsilon$ for $l=m$) \cite{Traytak95}.  We emphasize
that this is the exact solution (see further extensions in
\cite{Grebenkov19}).  A truncation of the matrix $\M$ to a finite size
$N \times N$ yields an accurate approximation for $P$ and thus for
$\kappa$, whose error decreases very rapidly as $N\to \infty$.  Note
that the ratio $P$ can be interpreted as the probability of reaction
on the patch (before escaping to infinity).

The knowledge of the correct asymptotic behavior (\ref{eq:asymptotic})
and the possibility of an accurate computation of $\kappa(\sigma)$ via
Eq. (\ref{eq:Traytak}) allow us to remediate the empirical formula
(\ref{eq:Dagdug}).  For this purpose, we keep the structure of this
formula but replace numerical coefficients $2.6$ and $-3.2$ by
adjustable constants $A$ and $B$.  By imposing the asymptotic behavior
(\ref{eq:asymptotic}), we first relate $A$ and $B$ in the leading
order of the limit $\sigma \to 1$: $(2/\pi) (1 + A - B) = 3\pi/8$,
from which $B = 1 + A - 3\pi^2/16$.  Then fitting $(1-\sigma)^{3/2}
\kappa(\sigma)$ against the semi-analytical solution
(\ref{eq:Traytak}) over $\sigma$ from $0$ to $1$, we determine the
optimal value $A \approx 2.32$, from which $B \approx 1.47$.  We
arrive thus to another empirical approximation,
\begin{equation} \label{eq:DG}
\frac{R}{D} \kappa_{\rm new}(\sigma) = \frac{2}{\pi} \sqrt{\sigma} \frac{1 + 2.32\sqrt{\sigma} - 1.47 \sigma^2}{(1-\sigma)^{3/2}} \,,
\end{equation}
which correctly reproduces both limits $\sigma \to 0$ and $\sigma \to
1$.

We use the truncated semi-analytical solution (\ref{eq:Traytak}) as a
benchmark for illustrating the accuracy of both empirical
approximations.  To control the effect of the truncation order,
$\kappa$ was computed via Eq. (\ref{eq:Traytak}) for $N = 20$ and $N =
100$, and the maximal relative error between these two computations
was $0.2\%$ over the considered range of $\sigma$ from $0.01$ to
$0.99$ (note that larger truncation orders may be needed for smaller
$1-\sigma$).  Figure \ref{fig:Dagdug}(b) illustrates the behavior of
the effective reactivity $\kappa$ (rescaled by $R/D$) versus $\sigma$.
First of all, one can appreciate the correct form of the asymptotic
behavior (\ref{eq:asymptotic}) as $\sigma\to 1$.  For small $\sigma$,
the original approximation (\ref{eq:Dagdug}) is close to the
semi-analytical solution, as expected.  However, deviations start to
emerge for $\sigma \gtrsim 0.6$ and become dramatic as $\sigma\to 1$,
as discussed above.  Finally, the improved empirical approximation
(\ref{eq:DG}) is accurate over the entire range of $\sigma$, not
exceeding the relative error of $6\%$.  This is clearly illustrated on
Fig. \ref{fig:Dagdug}(c).

In summary, we inspected limitations of the empirical approximation
for the effective reactivity proposed in \cite{Dagdug16}.  In the
limit $\sigma\to 1$, this approximation captured correctly the
exponent of the diverging behavior, $\kappa \propto
(1-\sigma)^{-3/2}$, but failed to get the correct prefactor.  As a
consequence, the relative error of this approximation was exceedingly
high as $\sigma\to 1$.  Using recent asymptotic results
\cite{Martin20,Grebenkov24}, we amended this problem and proposed an
improved version (\ref{eq:DG}) of this approximation, which reproduces
correctly both limits $\sigma\to 0$ and $\sigma\to 1$.  Its accuracy
was validated by comparison with the semi-analytical solution
(\ref{eq:Traytak}).  Beyond its own interest, the effective reactivity
$\kappa$ gives access to several important quantities such as the
probability of reaction, $P = 1/(1 + D/(\kappa R))$, the capacitance
of the structured target with a reactive patch, ${\mathcal C} =
\kappa/(4\pi D)$, as well as its hydrodynamic Stokes drag
\cite{Dagdug16,Hubbard93}.  Our improved formula (\ref{eq:DG}) allow
one to access these quantities for the entire range of the covered
fraction $\sigma$ that would be difficult to achieve by other means.

\begin{acknowledgments}
The author thanks Dr. A. T. Skvortsov for fruitful discussions and
acknowledges the Alexander von Humboldt Foundation for support within
a Bessel Prize award.
\end{acknowledgments}

\section*{Data Availability Statement}

The data that support the findings of this study are available from
the corresponding author upon reasonable request.


\begin{thebibliography}{99}

\bibitem{Collins49}		F. C. Collins and G. E. Kimball,
				Diffusion-controlled reaction rates,
				J. Coll. Sci. {\bf 4}, 425 (1949).

\bibitem{Dagdug16}		L. Dagdug, M.-V. V\'azquez, A. M. Berezhkovskii, and V. Yu. Zitserman,
				Boundary homogenization for a sphere with an absorbing cap of arbitrary size,
				J. Chem. Phys. {\bf 145}, 214101 (2016).


\bibitem{Martin20}		P. A. Martin and A. T. Skvortsov, 
				Scattering by a sphere in a tube, and related problems, 
				J. Acoust. Soc. America {\bf 148}, 191-200 (2020).

\bibitem{Martin22}		P. A. Martin and A. T. Skvortsov, 
				On blockage coefficients: flow past a body in a pipe, 
				Proc. R. Soc. A {\bf 478}, 20210677 (2022).

\bibitem{Skvortsov23}		A. T. Skvortsov, L. Dagdug, A. M. Berezhkovskii, and S. M. Bezrukov, 
				Blockage coefficient of cylindrical blocker and diffusion resistance of membrane channels, 
				Phys. Fluids {\bf 35}, 011702 (2023).

\bibitem{Grebenkov24}		D. S. Grebenkov and A. T. Skvortsov, 
				Boundary Homogenization for Target Search Problems, 
				Chapter 11 in ``Target Search Problems'', Eds. D. S. Grebenkov, R. Metzler, and G. Oshanin 
				(Springer: Cham, Switzerland, 2024), pp. 247-279 [preprint is available as arXiv:2310.14322].


\bibitem{Dagdug17}		L. Dagdug, A. M. Berezhkovskii, V. Yu. Zitserman,
				Note: Effect of a small surface defect on the Smoluchowski rate constant and 
				capacitance of a spherical capacitor,
				J. Chem. Phys. {\bf 147}, 106101 (2017).


\bibitem{Traytak95}		S. D. Traytak,
				Diffusion-controlled reaction rate to an active site,
				Chem. Phys. {\bf 192}, 1-7 (1995).

\bibitem{Grebenkov19}		D. S. Grebenkov, 
				Spectral theory of imperfect diffusion-controlled reactions on heterogeneous catalytic surfaces, 
				J. Chem. Phys. {\bf 151}, 104108 (2019). 

\bibitem{Hubbard93}		J. Hubbard and J. F. Douglas,
				Hydrodynamic friction of arbitrarily shaped Brownian particle,
				Phys. Rev. E {\bf 47}, R2983(R) (1993).



\end{thebibliography}
\end{document}